# Data Encoding for VQC in Qiskit, A Comparison With Novel Hybrid Encoding

Hillol Biswas

Department of Electrical and
Computer Engineering, Democritus
University of Thrace, Xanthi 67100,
Greece

*Abstract*—If quantum machine learning emulates the ways of classical machine learning, data encoding in a quantum neural network is imperative for many reasons. One of the key ones is the complexity attributed to the data size depending upon the features and types, which is the essence of machine learning. While the standard various encoding techniques exist for quantum computing, hybrid one is not among many, though it tends to offer some distinct advantages, viz. efficient qubits utilization and increased entanglement, which fits well for variation quantum classifier algorithm by manipulating the essential criteria of ZZFeatureMaps and RealAmplitudes. While Amplitude encoding can turn traits normalized into quantum amplitudes, encoding an angle by using Ry gates to encode feature values into rotation angles, and phase encoding by using Rz gates to encode extra feature information as phase is plausible to combine all together. By combining these three methods, this paper demonstrates that efficient qubit usage is ensured as Amplitude encoding reduces the required qubits, Angle encoding makes state freedom better and is used for expressive encoding, and Phase-based distinction. Finally, using classical optimizers, the hybrid encoding technique through VQC is fit in training and testing using a synthetic dataset, and results have been compared to the standard VQC encoding in qiskit machine learning ecosystems.

Keywords—Quantum Machine Learning, Data Encoding, Quantum Neural Networks, Variational Quantum Classifier, IBM QPU

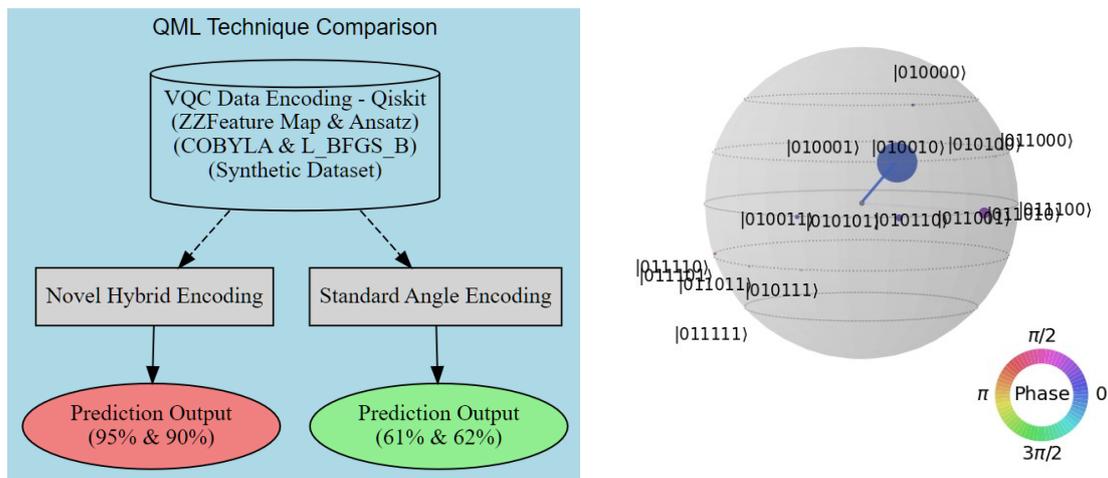

**Graphical Abstract**

## I. INTRODUCTION

Richard Feynman's 1982 vision of a quantum computer reads, "Therefore my question is, Can physics be simulated by a universal computer?"[1] had a novel beginning with Deutsch, who significantly advanced it in 1985. Although these points are up for debate, Deutsch's idea is generally regarded as the first blueprint for a QC since it is both sufficiently straightforward and specific to consider actual machines and adaptable enough to serve as a universal quantum simulator. Deutsch demonstrated that any unitary evolution could be created by forcing the 2-state systems to evolve through a limited number of basic operations; hence, the evolution could mimic that of any physical system, quantum simulator, albeit there is disagreement on both things. In essence, Deutsch's system resembles a register machine more than a Turing machine and is really a series of 2-state systems [2].

Deutsche, 1985 [3] in fact examined a few of the many links that exist between the quantum theory of computation and the rest of physics. The quantum complexity theory permits a concept of "complexity" or "knowledge" in a physical system that is more grounded in reality than traditional complexity theory. By creating quantum Turing machines, Deutsch expanded on Feynman's idea and proposed that if quantum computers could outperform digital computers at solving quantum mechanical issues, they would also be able to outperform them at solving classical ones [4]. First for quantum Turing machines, then for quantum circuits [5], Deutsch was the first to establish a formal model of quantum computation [6]. In the follow-up, Grover search algorithm [7], Shor algorithm for prime factorization [8], Deutsche-Jozsa for rapid solution [9], Berstein Vazirani algorithm for complexity problem [10] transpired, and many others recently augmented the pursuit of quantum computing research, gradually branching out to other domains, including quantum machine learning (QML). Developing quantum circuits and gates for performing quantum computation is the crux of solving complex problems in this endeavour by various quantum mechanics principles and mathematical techniques [11].

The fundamental unit of information in quantum computing, known as a qubit, can exist in one of three states: 0, 1, or a superposition of both, as demonstrated by a linear combination of 0 and 1. A collection of numbers, a formula for creating them, and an algebra that acts on numbers and digits using a universal set of operators produce Quantum computing techniques using many qubits. The qubit is the fundamental building block of quantum computation, with $|\psi\rangle = |0\rangle + |1\rangle$ (where, C and $|0\rangle$, $|1\rangle$ The square of the amplitudes is the probability to measure in the two-dimensional Hilbert space H2. In quantum dynamics, the probability conservation property $|\alpha\rangle^2 + |\beta\rangle^2 = 1$ is always maintained, regardless of whether the qubit is in the 0 or 1 state. One could argue that the discipline of quantum machine learning began with the HHL algorithm [12], proposed in 2008 by Aram Harrow, Avinatan Hassidim, and Seth Lloyd to solve a linear equation on a quantum computer. The recent interest and research in quantum computing also led to branching out in quantum machine learning, translating the approach of the classical algorithms [13]. Observation of spontaneous activity or phenomenon can inspire formulation of a quantum algorithm viz. QANA inspired from migratory bird flight path [14]. It is no wonder that a quantum circuit-based model reciprocatively leads to quantum machine learning for finding out scope of observing large-scale patterns. Thus, Quantum computing techniques for developing circuits invariably lead to quantum machine learning (QML) when extended to significant data size and complexity.

However, due to high resource costs, developing workable solutions for noisy intermediate-scale quantum (NISQ) devices is a crucial top priority. Loading classical data into quantum machine learning algorithms, nonetheless, is a persistent issue. Representing the data as a quantum state is a standard method that can be accomplished in several ways [15]. Theoretically, faster quantum algorithms frequently presume effective data loading, which is easier said than done. Generally speaking, (i) the data encoding, which specifies how the quantum computer's state represents the data, and (ii) the data itself determines the loading routine's runtime complexity. In many instances, loading the data takes at least exponentially long, negating any possible speedup [16]. Regardless of the learning process and type, the best machine learning algorithms use the fewest resources and have the lowest task-related error rate (as demonstrated by incorrect input categorization, subpar clustering, and low strategy reward). Complex tasks are finding the parameters and starting values that result in the best solution or developing strategies that lower the algorithm's complexity class [17]. Three primary challenges in quantum computation have emerged in recent decades: how to encode classical data into a quantum computer, process quantum data, and retrieve valuable information from processed quantum data [18]. However, many intriguing topics remain when examining the impact of data encoding in QML. First, is it possible to comprehend and measure the generalization capacity using the framework created here? of quantum models and thereby provide helpful guidance for model selection [19]?

## II. RELATED WORKS

Loading data into a quantum computer is essential for conducting an algorithm. Almost all algorithms that process input data start with this task. Following the first loading procedure, a particular encoding represents the data by qubits. Before processing the input, each method assumes that a specific data encoding is utilised for computations. Regretfully, efficient data loading is not always possible. In the worst situation, loading takes an exponential amount of time. This slows down algorithms with an otherwise linear or logarithmic runtime: They have an exponential runtime and an exponential loading time. This negates the potential for an algorithm to speed up linearly or exponentially, which was one of the original motivations for using a quantum computer. This entails recognizing the patterns for encoding data in quantum computations. Every pattern explains how input data is loaded at the start of a quantum algorithm in a particular encoding. Each encoding pattern is refined since data is loaded during the INITIALIZATION stage. From the most straightforward encoding i.e., basis encoding, then quantum associated memory and further amplitude encoding had been discussed by Weigold et al [20].

Further, the method of encoding data is also known as applying a quantum feature map ϕ [21] in the context of quantum machine learning. The authors note the connection between machine learning kernels and quantum feature maps. Each feature map implicitly defines a quantum kernel. Once the quantum feature map has been applied, the data can be loaded and evaluated in a high-dimensional space. As a result, every encoding that they discussed in the study specifies a quantum kernel and produces a quantum feature map ϕ [16]. Three encoding patterns, including binary encoding, comparison-based unary encoding, and comparison-based one hot encoding, their effects on scalability, and their usability, to discuss the merits and disadvantages of each, Gogeiβl et al. use the entertaining (but computationally demanding) Sudoku problem and its reduction to graph colouring as an example [22].

To put a concept into practice, explain the Schmidt decomposition process and how to use singular value decomposition to determine a quantum vector's Schmidt basis and Schmidt coefficients, Kumar and Ghosh describe how to build a quantum circuit graphically. As an example, they encode a set of 16 classical data into a 4-qubit quantum state [18] which elucidates the data encoding complexity.

Time-efficient techniques for C2Q data encoding and Q2C data decoding are suggested by Mahmud et al. Compared to other approaches, they provide quantum circuits for C2Q with circuit depths lowered by a factor of 2. A time-efficient technique for decoding Q2C data is put forth, which involves employing the Quantum Wavelet Transform (QWT) to sample the output state.

The associated optimised quantum circuits are also shown. The IBM QASM simulator is used to simulate and assess the suggested C2Q and Q2C circuits and procedures [23].

The various data encoding schemes—Basis Encoding, Amplitude Encoding, Angle Encoding, Dense Angle Encoding, General Qubit Encoding, Wavefunction Encoding, and Hamiltonian Encoding—are described to address the different types of data encoding techniques used for solving particular research problems in quantum machine learning [24].

Rath and Date investigated several classical-to-quantum mapping techniques for encoding classical data, including basis encoding, angle encoding, and amplitude encoding. They carried out an comprehensive empirical study that includes ensemble techniques like Random Forest, LightGBM, AdaBoost, and CatBoost and well-known machine learning algorithms like K-Nearest Neighbours, Support Vector Machines, and Logistic Regression. The results show that quantum data embedding improves F1 scores and classification accuracy, which is especially significant for models that naturally gain from better feature representation. It was noticed that there were subtle variations in running time, with more computationally demanding models showing noticeable shifts and low-complexity models showing modest increases [25].

Every work mentioned or summarised in it contributes significantly to the QC conversation, but it always has some significant flaws and its own set of restrictions. These range from not giving a thorough rundown of the principal quantum encoding methods to not thoroughly examining the mathematical foundations, necessary qubit allocations, computing effectiveness, and possible real-world uses. Our survey, on the other hand, fills these gaps by providing a thorough comparative analysis of the most well-known encoding schemes, with a focus on their mathematical formulations, qubit requirements, runtime complexities, and practical applicability—all of which have not received enough attention in the body of existing literature [26]. In quantum machine learning, the use of a variational quantum classifier (VQC), in qiskit, conceptually related to the variational quantum eigensolver (VQE) though serving a different purpose, has been cited as a quantum neural network application for sensor data binary classification [27].

The data encoding issue restricts the scope of benchmarking tasks that may currently be implemented experimentally in addition to acting as a bottleneck on the potential speedup in many supervised QML methods [28]. The number of features in High Energy Physics (HEP) datasets is often higher than the order of the number of qubits currently accessible, making them high-dimensional. This makes it difficult for data encoding circuits on existing quantum devices to process and receive direct input. A set of reduced features is commonly utilised as an input to the QML models to overcome this difficulty [29]. Hence, dataset size is paramount among other things in machine learning. So, as the complexity scenario in the quantum applications corresponds, The authors [25] using a dataset that includes 7043 unique consumers and 20 attributes, including a binary target field that indicates turnover accessible on the Kaggle website, chose to keep the minority classes and undersample the majority classes, yielding 3738 records overall, because of the simulator's difficulties in processing vast amounts of data. In error analysis, the number of qubits and epochs used in the circuit and training, and the corresponding hyperparameter tuning are essential part to consider for the application [30]Therefore, based on the above related works, a research gap has been identified in data encoding details on quantum neural networks and their corresponding implications in Quantum machine learning, which involves a significant amount of data with any possible hybrid technique.

The paper compares a few existing encoding methods and their corresponding quantum circuit visualization for machine learning through QNN in qiskit implementation in IBM's quantum processing ecosystem [20], which is the contribution. The novelty of this work is in relating data encoding, examining a hybrid encoding technique, and using VQC-based classification to understand the inherent components in inspecting generalization issues in adopting QML.

### III. Data Encoding For Quantum Neural Network

The advantage of amplitude encoding is that it requires fewer qubits to encode. For example, suppose a dataset X contains m samples with each n-feature. Thus, one amplitude vector whose length is nm is needed to encode dataset X. A quantum system of N-qubit generates 2n amplitude. So, it requires $N \geq \log nm$ qubits to encode n-dimensional data with m samples. The disadvantage is that the circuit for amplitude encoding is intense because amplitude encoding encodes the entire dataset at a time, increasing the circuit's depth [30].

A study highlights how quantum data embedding might improve traditional machine learning models and stresses the significance of balancing computational costs and performance gains. Future studies should focus on enhancing quantum encoding procedures to maximise computational effectiveness and investigate scalability for practical uses. Nonetheless, the encoding methods comprise basis encoding, superimposition encoding, angle encoding, and amplitude encoding using Pennylane [25].

Processing images on quantum computers requires representing image data as quantum states [28], whereas existing techniques are apart from amplitude encoding, NEQR [31], and FRQI [32]. It was mentioned that choosing a particular data encoding technique for high-resolution satellite images is an intriguing study topic using satellite data from synthetic aperture radio (SAR). When selecting data encoding algorithms, the characteristics of the data and the intended remote sensing application must be considered. Another factor to consider while choosing such a data encoding scheme is the enormous size of photographs. The first stage in processing high-resolution satellite photos is encoding image pixel data. The three standard techniques of basis, angle, and amplitude encoding provide the foundation of embedding methods for data encoding. Using amplitude encoding, they proposed four different types of quantum compression techniques (QCT) for image data processing [33].

From the standpoint of machine learning, QNNs are once more algorithmic models that, like their classical counterparts, may be trained to uncover hidden patterns in data. These models can use trainable weights to parameterize quantum gates to process classical data (inputs) after loading it into a quantum state. A generic QNN example includes the data loading and processing processes. The weights can then be trained via backpropagation by plugging the output from measuring this state into a loss function. From a quantum computing standpoint, QNNs are quantum algorithms that may be taught variationally using classical optimizers. They are built on parametrised quantum circuits. These circuits include an ansatz (with trainable weights) and a feature map (with input parameters) [34]. A practical implementation of a variational quantum classifier. The measured bitstrings are interpreted as the classifier's output in the variational quantum classifier (VQC) [35] which has also been used for Quantum LSTM for natural language processing (NLP) representing each gate [30]. This variational method instantiates a neural network classifier by building a quantum circuit and matching a neural network. Some of the different ways to supply labels are a numpy array with categorical string labels, a one-dimensional numpy array with numeric labels such as [0, 1, 2, …], or plain labels [35]. However, how can the data encoded into the QNN be visualized from qiskit ecosystems for the ZZFeatureMaps and Ansatz, i.e., RealAmplitudes? In classical or quantum machine learning problems, the data are inevitably of significant size and variables, which is the crux of using them to find patterns. Then, how are the data frame elements encoded in the algorithms?

A quantum neural network (QNN) in qiskit demands data encoding in such a way that the classical optimizers, viz. COBYLA [36], Limited-memory BFGS Bound optimizer [37] can act to minimize the value of differentiable functions.

This relatively complex technique encodes classical data into quantum states via the angle. Since it only encodes a single data point at a time rather than a whole dataset, it varies from the other two encoding techniques [30]. Considering the above, the research community has extensively cited amplitude encoding as a relatively simple technique.

*A. Dataset Selection*

Synthetic data was generated using SKLEARN make_classification, comprising six variables and a class of binary variables 0 and 1 with random state = 42, shuffle = False, and without any imbalance criteria. The dataset consists of 1000 samples created using random functions in the library. With sample data from this set, the following different type of data encoding is demonstrated. Standard scaling with sklearn library was applied for the variables.

*B. Amplitude encoding*

Amplitude encoding, Fig. 1 and 2, is a way to store quantum data that uses quantum superposition to show regular data. Amplitude encoding stores data in the probability amplitudes of a quantum state, while angle encoding uses rotation gates.

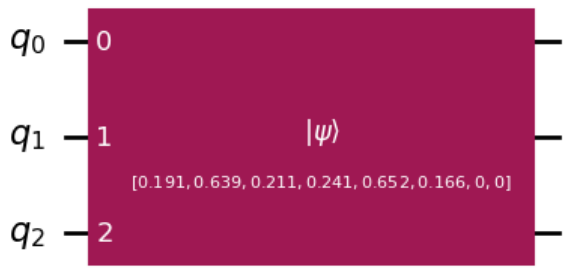

Fig. 1. Amplitude encoding in 3 qubits

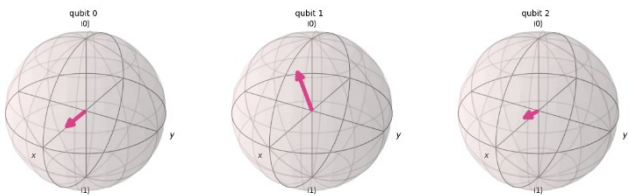

Fig. 2. Qsphere of 3 qubits for amplitude encoding

The amplitude encoding circuit represents the sample point [0] using the norm mathematical formula, and the state ψ is described below. It means a complex vector of 8 dimensions in a 3-qubit system ($2^3$ = 8). Each coefficient represents amplitude, and squaring it provides the probability of the measuring state. The amplitude represented by zero implies the absence of those states in the measurement.

$$|\psi\rangle = [0.191\ 0.639\ 0.211\ 0.241\ 0.652\ 0.166\ 0\ 0]$$

Further entanglement analysis reveals that the purity is 0.658, indicating that it is less than 1 in a mixed state with a Von Neuman entropy of 0.758. The higher the entropy, the more entanglement results. Schimdt coefficients represent the degree of entanglement, and the presence of multiple coefficients implies the states in a mix state.

## C. Basis Encoding

In the basis encoding, Fig. 3 and 4, six qubits were created for six corresponding variables, and the sample point [0] thresholding to binary {0,1} and applying X gate at 1, state basis vector has been created for the quantum circuit.

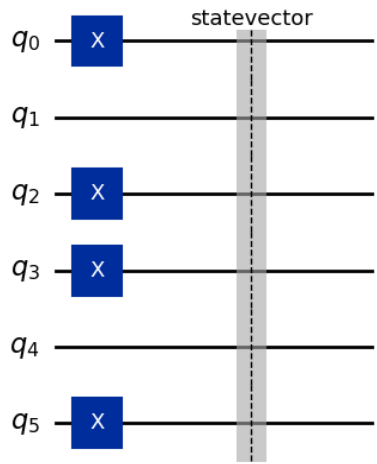

Fig. 3. Basis encoding in 6 qubits

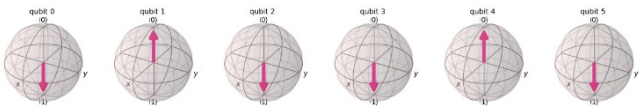

Fig. 4. Bloch sphere in 6 qubits for basis encoding

## D. Angle Encoding

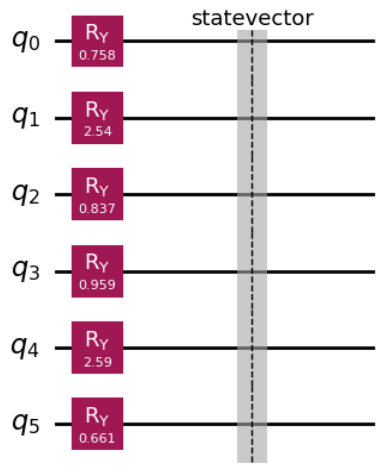

Fig. 5. Angle encoding involving 6 qubits

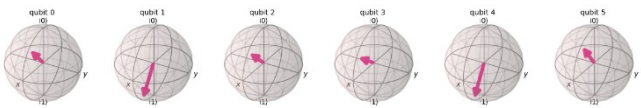

Fig. 6. Bloch sphere in 6 qubits for angle encoding

Angle encoding is a way to store quantum features, Fig. 5 and 6. It uses rotational gates to map traditional data points to quantum states. One of the easiest ways to store data in quantum devices is in this way. The state array_to_latex yields for the state as
$0.0572475166|000000\rangle+0.0227939382|000001\rangle+0.1851837702|000010\rangle+0.0737336334|000011\rangle+0.0254608809|000100\rangle+0.010137623|000101\rangle+0.0823606366|000110\rangle+210021916\cdot3437191\cdot5437425\cdot7437360$

|000111⟩+0.0297723912|001000⟩+0.0118543141|001001⟩+0.0963074729|001010⟩+0.0383462325|001011⟩+0.0132412959|001100⟩+0.0052722161|001101⟩+0.0428328291|001110⟩+0.0170545189|001111⟩+0.2024557566|010000⟩+0.0806107281|010001⟩+0.6549021249|010010⟩+0.2607588838|010011⟩+0.0900423669|010100⟩+0.0358516887|010101⟩+0.2912682674|010110⟩+0.1159727315|010111⟩+0.1052900169|011000⟩+0.0419227641|011001⟩+0.3405912331|011010⟩+0.1356113935|011011⟩+0.046827823|011100⟩+0.0186451844|011101⟩+0.1514782356|011110⟩+0.0603132806|011111⟩+0.0196441865|100000⟩+0.0078216209|100001⟩+0.0635448442|100010⟩+0.0253013115|100011⟩+0.0087367684|100100⟩+0.0034786724|100101⟩+0.0282616226|100110⟩+0.0112527794|100111⟩+0.0102162407|101000⟩+0.004067746|101001⟩+0.0330474066|101010⟩+0.0131583095|101011⟩+0.0045436816|101100⟩+0.0018091334|101101⟩+0.0146978617|101110⟩+0.0058521692|101111⟩+0.0694716361|110000⟩+0.0276611506|110001⟩+0.2247262459|110010⟩+0.0894780499|110011⟩+0.0308975682|110100⟩+0.0123023198|110101⟩+2452421363·342175·5421276·7421292|110110⟩+0.0397954375|110111⟩+0.0361297197|111000⟩+0.0143855777|111001⟩+0.1168721039|111010⟩+0.0465343419|111011⟩+0.016068723|111100⟩+0.0063979977|111101⟩+0.0519789659|111110⟩+0.0206961874|111111⟩

for the corresponding six qubits. Similarly, as in the explanation of amplitude encoding states entanglement, the corresponding probabiity amplitudes and probabilities leading to entanglement can also be explained in six qubits. |010010⟩ state appears with largest amplitude of 0.654 with corresponding probability of 0.42 for measurement.

*E. Phase Encoding*

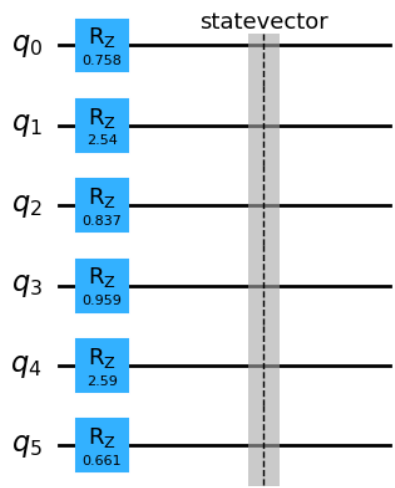

Fig. 7. Phase encoding in 6 qubits

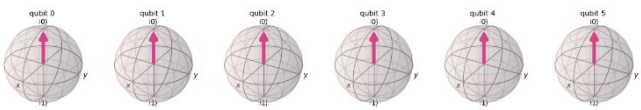

Fig. 8. Bloch sphere in 6 qubits

Phase encoding, Fig. 7 and 8, is a way to store quantum information. It maps classical data points to quantum states by changing the phase of computational basis states with θ $w$ z rotations.

*F. Hybrid Encoding*

A hybrid encoding function is created with a conditional relation of amplitude encoding if the power of 2 matches the dataset's length. Otherwise, angle and phase encoding with Ry and Rz rotations is applicable. Enough qubits are ensured for all encoding types matching the number of features with the length of the dataset. The state array to latex of the Hybrid encoding for six qubits, Fig. 9 for a sample first data generated as:

$(0.0249476376−0.028409179i)|010000⟩+(0.0025667674−0.001521704i)|010001⟩+(0.5105389908−0.8095017615i)|010010⟩+(0.0581060718−0.0482589257i)|010011⟩+(0.0052418374−0.0014931241i)|010100⟩+(0.0004298537+1.61444·10^{−5}i)|010101⟩+(0.1250916854−0.0581947923i)|010110⟩+(0.0108098288−0.0013078054i)|010111⟩+(0.009290638+0.0011240092i)|011000⟩+(0.0006696695+0.0003115417i)|011001⟩+(0.2367238995−0.0088908486i)|011010⟩+(0.0179808531+0.0051218005i)|011011⟩+(0.0010378214+0.0008619434i)|011100⟩+(5.67983·10^{−5}+9.00584·10^{−5}i)|011101⟩+(0.0293753001+0.0174151004i)|011110⟩+(0.0017784046+0.0020251623i)|011111⟩$.

As mentioned above, the hybrid encoding entails 16 nonzero terms culminating into a six-qubit superposition. State $|010010\rangle$ has a probability of 0.868, indicating a possible collapse in this state upon measurement.

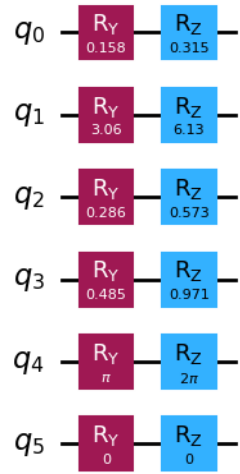

Fig. 9. Quantum circuit of six qubits for the hybrid encoding

The choice of Feature Map determines the kind of quantum encoding employed in Variational Quantum Classifiers (VQC). VQC usually uses parameterised rotation or angle encoding. In VQC Angle Encoding (Parameterised Rotation Encoding) is with an encoding Type ZZFeatureMap, often utilised in VQC. It uses rotations (such as $Ry(xi)$ R y (x i) or $Rz(xi)$ R z (x i) gates) to translate classical data $xi$x i into quantum states. For instance, ZZFeatureMap uses CX (controlled-X) gates to entangle qubits with RZ gates. The benefit is envisaged as qubit-efficient because a single qubit can store several characteristics using different angles. Nonetheless, the standard VQC encoding uses angle encoding type.

Fig. 10 depicts the six qubits of the Bloch sphere, wherein the red arrow points toward the corresponding state of each one. In qiskit, VQC usually requires ZZFeatureMap and Ansatz i.e., RealAmplitudes corresponding to it, which is the training parameters for the compilation. However, hybrid encoding was ensured instead of ZZFeaturesMap, a hybrid quantum circuit, Fig. 9, which is comprised of angle and phase encoding matching with the dataset number of variables treated as features. Fig. 11 depicts the decomposed hybrid circuit with two layers of Hadamard Gates, and Fig. 12 depicts the decomposed ansatz with rotational and control-NOT (C-NOT) gates, respectively. The Bloch spheres depict the state vectors in different orientations in Fig. 2, 4, 6, 8 and 10 for the corresponding encoding techniques.

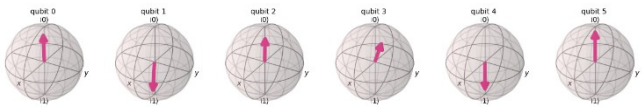

Fig. 10. Bloch sphere of six qubits for the hybrid encoding

*G. VQC Run and with Hybrid encoding*

The decomposed circuit, Fig. 11, depicts the VQC encoding technique with a hybrid feature map. Gates of Hadamard (H, in red), applied initially to all qubits, enable quantum interference by generating superposition. Gates for Phase Rotation (P, in Blue), the parameters of these gates rely on the classical inputs ($x_i$). This demonstrates that the circuit uses rotational parameters to encode data. Blue dots and lines indicate controlled-Z (CZ) gates introducing qubit-to-qubit paired entanglement. ZZFeatureMap has a feature that improves feature correlations. Once interaction terms are introduced, the Second Hadamard layer reverses the initial Hadamard effects.

Fig. 12 depicts a paremeterized quantum circuit comprising layers of Rotation (RY and RX Gates) in the corresponding ansatz. To add trainable angles, each qubit receives parameterized RX and RY gates. The CNOT Entanglement Layer (CX) increases expressibility by generating entanglement between qubits. Several Repetitions (parameter for reps) control depth: more trainable parameters result from more repetitions. The 18 number of trainable parameters corresponding to $R_y$ rotation. The two layer shallow architecture appears efficient for noisy intermediate state quantum (NISQ) devices.

Fig. 13 and 14 depict the standard VQC circuit encoding with a feature map and real amplitude, i.e., ansatz, and the decomposed circuit, respectively. The circuit depth, width, number of parameters, and counts of operations are 2, 6, 24, OrderedDict([('ZZFeatureMap', 1), ('RealAmplitudes', 1)] respectively. Qiskit reveals, for the vqc with hybrid encoding, the numbers are 2, 6, 0, OrderedDict([('ry', 6), ('rz', 6)], respectively. In contrast, the decomposed vqc with standard encoding yields the same as 35, 6, 24, OrderedDict([('cx', 30), ('p', 22), ('ry', 18), ('h', 12)]) respectively revealing the increase in depth of the circuit.

After the usual scaling of the data with the MimMax scaler, albeit in the range of 0 to π, the data is fed in the VQC with a hybrid feature map and corresponding ansatz of equal qubits of the number of variables. The VQC is then fit to the dataset using the simulator/IBM quantum processing unit IBM_Sherbrooke, a 127-qubit real quantum computer, which they make available for an IBMID through access and generate a corresponding token. Loss history, parameter history, accuracy, and confusion matrices are stored during the training process of 100 iterations each for the classical optimizers COBYLA, [36] and L_BFGS [37]. The results are displayed in the following section and subsequent Figures. The classical SKLEARN library is used for the train and test dataset split in the 80:20 ratio with a random state of 42. A callback function is used to track the fit history, and 100 iterations have been used with classical optimizers COBYLA and L_BFGS_B for comparison purposes.

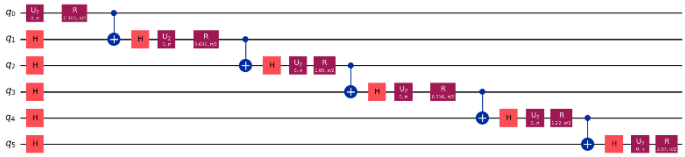

Fig. 11. Decomposed hybrid feature map

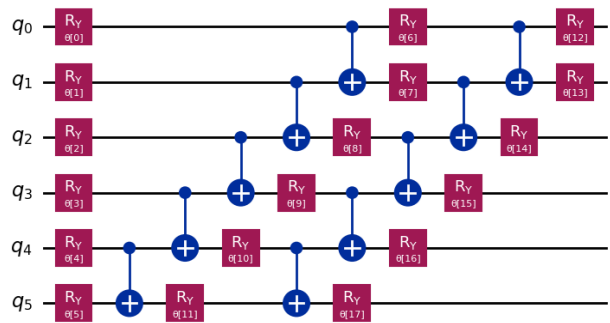

Fig. 12. *Decomposed ansatz corresponding to the hybrid feature map*

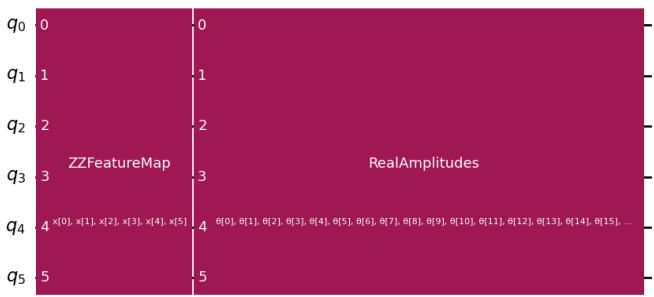

Fig. 13. Standard VQC quantum circuit with ZZfeatureMap and Ansatz

Fig. 13 depicts the hybrid VQC data, including six qubits for siz variables x[0] to x[5] in the ZZFeatureMap. The corresponding Ansatz, i.e., RealAmplitudes yields θ[i] for training parameters. The decomposed circuit, Fig. 14, depicts a three-layer architecture comprising feature encoding, a parametrized layer, and ansatz, which is the trainable part of the structure.

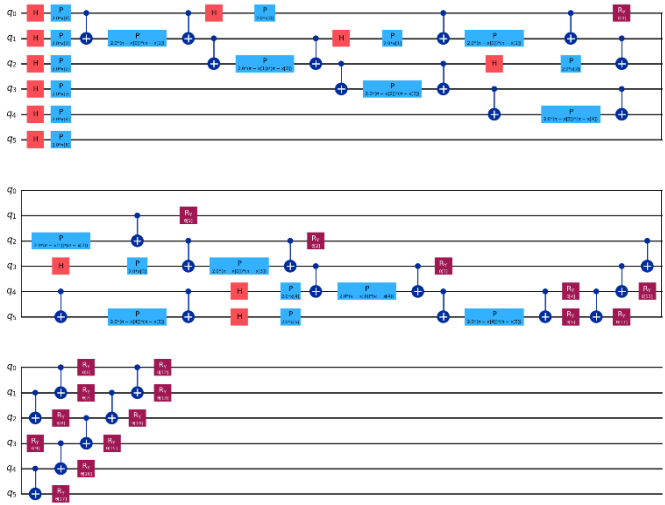

Fig. 14. Decomposed Standard VQC quantum circuit with ZZfeatureMap and Ansatz

## IV. VQC QISKIT RESULTS VISUALIZATION

Subsequent details and Figs. 13, 14, and 15 depict the VQC run for training after compiling the circuit with feature map and ansatz results.

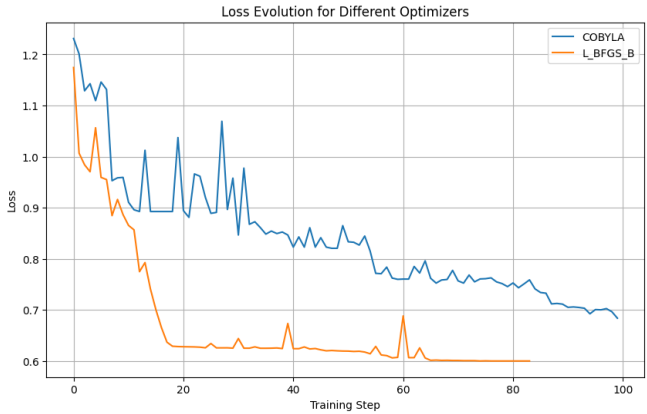

Fig. 15. Loss evolution with COBYla and L_BFGS_B classical optimizers for hybrid encoding

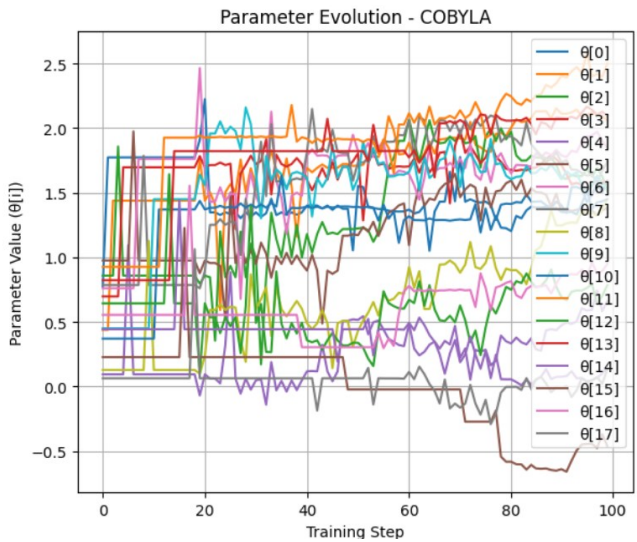

Fig. 16. Ansatz parameter evolution with optimizer COBYLA with VQC hybrid encoding

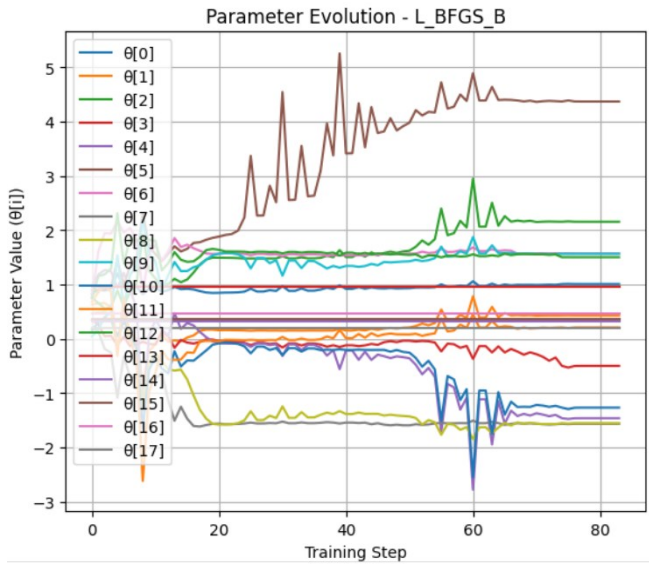

Fig. 17. Ansatz parameter evolution with optimizer L_BFGS_B with VQC hybrid encoding

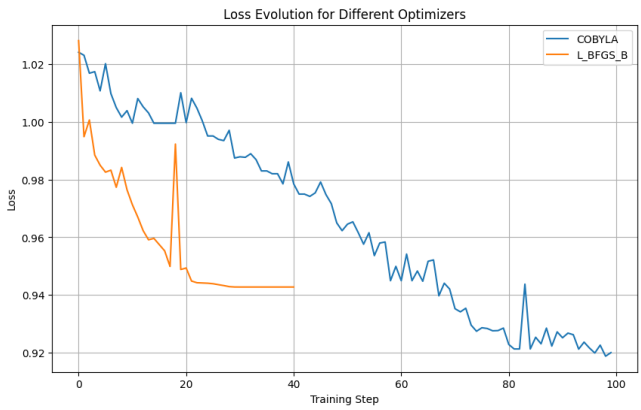

Fig. 18. Loss evolution with COBYla and L_BFGS_B classical optimizers for standard VQC encoding

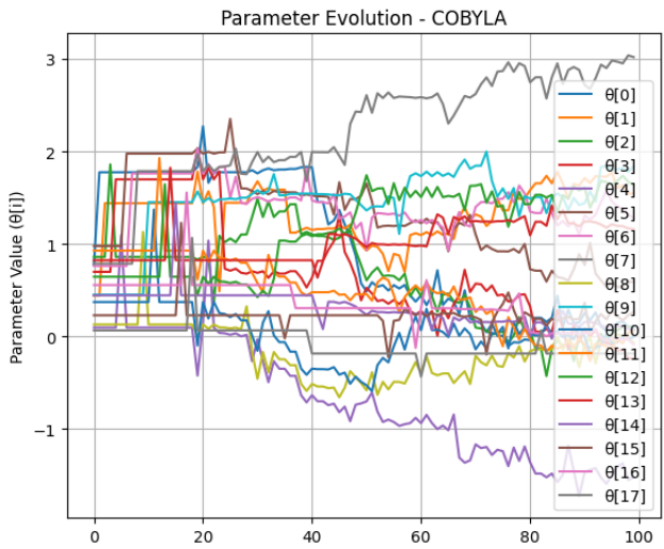

Fig. 19. Ansatz parameter evolution with optimizer COBYLA with VQC standard encoding

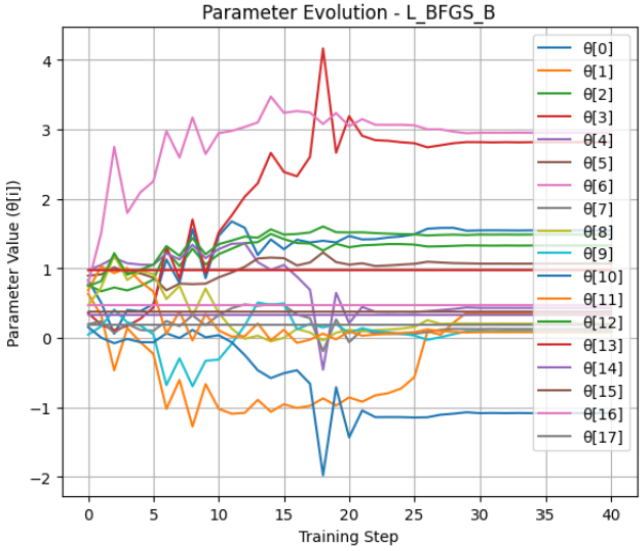

Fig. 20. Ansatz parameter evolution with optimizer L_BFGS_B with VQC standard encoding

After successful training with 100 iterations, for VQC with hybrid encoding and standard encoding, an accuracy of 0.95 and 0.90, compared with 0.61 and 0.62, was yielded for the respective run with optimizers COBYLA and L_BFGS_B., Figs. 15 to 20, respectively, depict the final trained parameters, totaling 18 numbers, each corresponding to 6 features input in qubits as a hybrid feature map. As Fig. 18 reveals while the COBYLA optimizer enables slightly better convergence with many iterations, with L_BFGS_B the convergence tends to reach early nearly half of the former.

## V. DISCUSSION AND CONCLUSION

The optimum use of circuit depth, width, and number of gate operations is crucial in the implementation in qiskit [38], IBM's SDK depends on the problem type. While the relation of the data encoding scheme with the number of qubits required and the runtime of preparing of quantum states for MxN data points and features are worth considering [24]. However, having an idea and visualization of a QNN circuit in qiskit in decomposed form and drilling down the components of ZZFeatureMap and RealAmplitudes can provide significant insight into how the circuits are configured for hardware efficient use and provide the scope of the corresponding architecture for refinement. Only as many qubits are used as features in an efficient representation; feature interactions are improved by quantum entanglement. As VQC, which originally stood for Variational Quantum Classifier for classification purposes, and it corroborates with RealAmplitudes Ansatz, which must be compatible with the ZZFeatureMap.

TABLE I. TRAINING PARAMETERS COMPARISON OF OPTIMIZERS

| COBYLA in Hybrid encoding | L_BFGS_B in Hybrid encoding | COBYLA in standard encoding | L_BFGS_B in standard encoding |
|---|---|---|---|
| θ[0] = 1.5946 | θ[0] = 1.0072 | θ[0] = 0.1086 | θ[0] = 1.5451 |
| θ[1] = 2.0606 | θ[1] = 0.4278 | θ[1] = 1.5445 | θ[1] = 0.3469 |
| θ[2] = 1.5157 | θ[2] = 2.1551 | θ[2] = 0.0722 | θ[2] = 1.3269 |
| θ[3] = 1.5906 | θ[3] = -0.5000 | θ[3] = -0.2439 | θ[3] = 2.8142 |
| θ[4] = 0.6612 | θ[4] = -1.4641 | θ[4] = -1.5249 | θ[4] = 0.4288 |
| θ[5] = 1.3816 | θ[5] = 4.3704 | θ[5] = 0.5923 | θ[5] = 1.0658 |
| θ[6] = 1.8456 | θ[6] = 1.5697 | θ[6] = 1.2793 | θ[6] = 2.9494 |
| θ[7] = 1.7393 | θ[7] = -1.5670 | θ[7] = 3.0189 | θ[7] = 0.1258 |
| θ[8] = 1.4480 | θ[8] = -1.5564 | θ[8] = -0.0664 | θ[8] = 0.2035 |
| θ[9] = 1.5385 | θ[9] = 1.5653 | θ[9] = 1.8399 | θ[9] = 0.0917 |
| θ[10] = 1.4506 | θ[10] = -1.2698 | θ[10] = -0.0085 | θ[10] = -1.0856 |
| θ[11] = 2.5006 | θ[11] = 0.2071 | θ[11] = 0.0565 | θ[11] = 0.0801 |
| θ[12] = 0.7850 | θ[12] = 1.5020 | θ[12] = 1.6343 | θ[12] = 1.4853 |
| θ[13] = 2.0960 | θ[13] = 0.9675 | θ[13] = 1.1610 | θ[13] = 0.9675 |
| θ[14] = 0.1475 | θ[14] = 0.3258 | θ[14] = 0.0745 | θ[14] = 0.3258 |
| θ[15] = -0.4612 | θ[15] = 0.3705 | θ[15] = 0.2613 | θ[15] = 0.3705 |
| θ[16] = 0.9485 | θ[16] = 0.4696 | θ[16] = -0.0890 | θ[16] = 0.4696 |

| θ[17] = 0.1332 | θ[17] = 0.1895 | θ[17] = -0.1862 | θ[17] = 0.1895 |

Therefore, in the light of above, this study entails discussing various data encoding techniques and corresponding customized a novel hybrid vis-vis-standard VQC encoding for a synthetic dataset. Using hybrid and standard angle encoding, the dataset is subsequently trained for significant iterations with two classical optimizers. The corresponding training records are well-preserved for visualization and subsequent inspection purposes for comparison in view of gaining meaningful insight. Table I depicts the respective optimizers' training parameters obtained after 100 iterations for both VQC with hybrid encoding and standard encoding. While a direct comparison is unlikely to yield a better candidate, Fig. 13 reveals a better convergence with the L_BFGS_B optimizer than COBYLA. As evident from above, a brief comparison of different standard data encoding methods in quantum computing through qiskit is a basis for proposing a novel hybrid encoding technique involving amplitude, phase, and angle encoding depending upon the dataset length. Based on the hybrid and standard encoding, VQC is trained and tested for a synthetic dataset of 1000 samples with two different classical optimizers. The training data records and appropriate accuracy matrices are presented for comparison purposes. As evident from the results, the accuracy score for hybrid encoding outperformed that of VQC with standard encoding. The basic idea of observing the training process in QML dependency with the number of variables and dataset length is directly related to the data encoding method in the quantum circuit. As quantum machine learning branches out from quantum computing and deals with significant datasets, this hybrid encoding technique further explores real-world datasets. Dimensionality reduction through principal component analysis (PCA) might appear a choice not to ignore; however, addressing the broad complexity of QML for speed-up, data encoding has been envisaged as the study's primary aim, hence did not transpire here. However, in the current pervasive advent of classical machine learning, it is essential to rigorously research and test corresponding quantum machine learning algorithms in whatever architecture appears promising for a possible candidate for future alternatives. The work must be extended using real-world and relatively more enormous datasets to examine the application generalization issue. Performance accuracy vis-à-vis computation time has not been compared, which is another limitation of this work. Further, the study used two classical optimizers, which can be extended with similar ones viz. SPSS, ADAM, or Gradient Descents. Moreover, developing a variant of hybrid VQC in conjunction with hybrid encoding with the possible use of quantum Fourier transform (QFT) [39], [40] for frequency-based feature transform and enhanced entanglement has been considered a future direction for further inspection of the generalization issue discussed in quantum neural networks-based quantum machine learning.

ACKNOWLEDGMENT

I thank the support and encouragement of everyone who enabled me to develop this paper.

REFERENCES

[1] R. P. Feynman, "Simulating Physics with Computers," vol. 21, pp. 467–488, 1982.
[2] A. Steane, "Quantum computing," no. October, 2000.
[3] D. Deutsch, "Quantum Theory, the Church-Turing Principle and the Universal Quantum Computer," *Proc. R. Soc. Lond. A. Math. Phys. Sci.*, vol. 400, no. No. 1818(Jul. 8, 1985), pp. 97–117, 1985, [Online]. Available: https://www.cs.princeton.edu/courses/archive/fall06/cos576/papers/deutsch85.pdf
[4] P. W. Shor, "Quantum Computing," vol. ICM, 1998.
[5] D. . . Deutsch, "Quantum Computational Networks," *Proc. R. Soc. London . Ser. A , Math. Phys. Publ. by R. Soc. Stable*, vol. 425, no. 1868, pp. 73–90, 1989, [Online]. Available: https://www.jstor.org/stable/2398494 Quantum computational networks
[6] D. Aharonov, "Quantum computation," pp. 1–78, 2024.
[7] L. K. Grover, "A fast quantum mechanical algorithm for database search," pp. 212–219, 1996, doi: 10.1145/237814.237866.
[8] P. W. Shor, "Polynomial-Time Algorithms for Prime Factorization and Discrete Logarithms on a Quantum Computer," vol. 41, no. 2, pp. 303–332, 1999.
[9] R. Deutsch, David; Jozsa, "Rapid solution of problems by quantum computation," pp. 553–558, 1992, [Online]. Available: https://royalsocietypublishing.org/doi/epdf/10.1098/rspa.1992.0167
[10] U. Bernstein, Ethan;, Vazirani, "Quantum Complexity general," 1993, doi: 10.1145/167088.167097.
[11] M. A. Nielsen and I. H. . Chuang, *Quantum Computation and Quantum Information*, 2010th ed. Cambridge University Press.
[12] A. W. Harrow, A. Hassidim, and S. Lloyd, "Quantum Algorithm for Linear Systems of Equations," *Phys. Rev. Lett.*, vol. 103, no. 15, p. 150502, Oct. 2009, doi: 10.1103/PhysRevLett.103.150502.
[13] F. Tacchino, "An artificial neuron implemented on an actual quantum processor," *npj Quantum Inf.*, no. November 2018, pp. 1–8, 2019, doi: 10.1038/s41534-019-0140-4.
[14] H. Zamani, M. H. Nadimi-shahraki, and A. H. Gandomi, "Engineering Applications of Artificial Intelligence QANA : Quantum-based avian navigation optimizer algorithm," *Eng. Appl. Artif. Intell.*, vol. 104, no. April, p. 104314, 2021, doi: 10.1016/j.engappai.2021.104314.
[15] R. Dilip, Y. Liu, A. Smith, F. Pollmann, and M. States, "Data compression for quantum machine learning," *Phys. Rev. Res.*, vol. 4, no. 4, p. 43007, 2022, doi: 10.1103/PhysRevResearch.4.043007.
[16] M. Weigold *et al.*, "Expanding Data Encoding Patterns For Quantum Algorithms," 2021.
[17] M. Schuld, I. Sinayskiy, F. Petruccione, S. Africa, and S. Africa, "An introduction to quantum machine learning," 2014.
[18] K. J. B. Ghosh, "Encoding classical data into a quantum computer," pp. 1–13, 2021.
[19] M. Schuld, R. Sweke, and J. J. Meyer, "Effect of data encoding on the expressive power of variational quantum-machine-learning models," *Phys. Rev. A*, vol. 103, no. 3, p. 32430, 2021, doi: 10.1103/PhysRevA.103.032430.
[20] M. Weigold, J. Barzen, F. Leymann, and M. Salm, "Data Encoding Patterns for Quantum Computing," pp. 1–11, 2020, doi: 10.5555/3511065.3511068.
[21] M. Schuld, N. Killoran, R. S. West, and T. Mv, "Quantum Machine Learning in Feature Hilbert Spaces," *Phys. Rev. Lett.*, vol. 122, no. 4, p. 40504, 2019, doi: 10.1103/PhysRevLett.122.040504.
[22] M. Gogeißl and W. Mauerer, "Quantum Data Encoding Patterns and their Consequences," pp. 27–37, 2024, doi: 10.1145/3665225.3665446.
[23] N. Mahmud *et al.*, "Efficient Data Encoding and Decoding for Quantum Computing," no. 1, pp. 765–768, 2022, doi: 10.1109/QCE53715.2022.00110.
[24] M. B. Pande, "A Comprehensive Review of Data Encoding Techniques for Quantum Machine Learning Problems," *2024 Second Int. Conf. Emerg. Trends Inf. Technol. Eng.*, pp. 1–7, 2024, doi: 10.1109/ic-ETITE58242.2024.10493306.
[25] M. Rath and H. Date, "Quantum data encoding : a comparative analysis of classical-to-quantum mapping techniques and their impact on machine learning accuracy," 2024, doi: 10.1140/epjqt/s40507-024-00285-3.
[26] M. A. Khan and M. N. Aman, "Beyond Bits : A Review of Quantum Embedding Techniques for Efficient Information Processing," *IEEE Access*, vol. 12, no. April, pp. 46118–46137, 2024, doi: 10.1109/ACCESS.2024.3382150.
[27] H. Biswas, "Sensor Signal Malicious Data Binary Classification :

A Comparison of QNN , and VQC," *2024 IEEE Pune Sect. Int. Conf.*, pp. 1–6, 2024, doi: 10.1109/PuneCon63413.2024.10894827.

[28] K. Shen, B. Jobst, E. Shishenina, and F. Pollmann, "Classification of the Fashion-MNIST Dataset on a Quantum Computer," pp. 1–15, 2024.

[29] V. Belis, P. Odagiu, and T. K. Aarrestad, "Machine learning for anomaly detection in particle physics," *Rev. Phys.*, vol. 12, p. 100091, 2024, doi: 10.1016/j.revip.2024.100091.

[30] S. Pandey, N. J. Basisth, T. Sachan, and N. Kumari, "Quantum machine learning for natural language processing application," *Physica A*, vol. 627, p. 129123, 2023, doi: 10.1016/j.physa.2023.129123.

[31] Y. Zhang, K. Lu, Y. Gao, and M. Wang, "NEQR: a novel enhanced quantum representation of digital images," *Quantum Inf. Process.*, vol. 12, no. 8, pp. 2833–2860, 2013, doi: 10.1007/s11128-013-0567-z.

[32] P. Q. Le, F. Dong, and K. Hirota, "A flexible representation of quantum images for polynomial preparation, image compression, and processing operations," *Quantum Inf. Process.*, vol. 10, no. 1, pp. 63–84, 2011, doi: 10.1007/s11128-010-0177-y.

[33] S. R. Majji, B. S. Manoj, and S. Member, "Sensor signal procesing Quantum Approach to Image Data Encoding and Compression," *IEEE Sensors Lett.*, vol. 7, no. 2, pp. 1–4, 2023, doi: 10.1109/LSENS.2023.3239749.

[34] Q. Ecosystem and Q. M. L. 0.8.2, "Quantum Neural Networks." [Online]. Available: https://qiskit-community.github.io/qiskit-machine-learning/tutorials/01_neural_networks.html

[35] Q. Ecosystem and Q. M. L. 0.8.2, "VQC." [Online]. Available: https://qiskit-community.github.io/qiskit-machine-learning/stubs/qiskit_machine_learning.algorithms.VQC.html

[36] "COBYLA | IBM Quantum Documentation." Accessed: Aug. 26, 2024. [Online]. Available: https://docs.quantum.ibm.com/api/qiskit/0.28/qiskit.algorithms.optimizers.COBYLA

[37] Q. Ecosystem and Q. A. 0.3.1, "L_BFGS_B." [Online]. Available: https://qiskit-community.github.io/qiskit-algorithms/stubs/qiskit_algorithms.optimizers.L_BFGS_B.html

[38] "Qiskit | IBM Quantum Computing." Accessed: Aug. 13, 2024. [Online]. Available: https://www.ibm.com/quantum/qiskit

[39] I. G. Karafyllidis, "Visualization of the Quantum Fourier Transform Using a Quantum Computer Simulator," vol. 2, no. 4, pp. 271–288, 2004.

[40] I. G. Karafyllidis, "Quantum Computer Simulator Based on the Circuit Model of Quantum Computation," vol. 52, no. 8, pp. 1590–1596, 2005.